\documentclass[nonacm]{acmart}

\usepackage{todonotes}      
\usepackage{enumitem}       
\usepackage{array,ragged2e}                 
\newcolumntype{L}[1]{>{\RaggedRight}p{#1}}  
\newcolumntype{R}[1]{>{\RaggedLeft}p{#1}}   
\newcolumntype{C}[1]{>{\Centering}p{#1}}    
\newcolumntype{M}[1]{>{\RaggedRight}m{#1}}  
\newcolumntype{B}[1]{>{\RaggedRight}b{#1}}  
\usepackage{soul}           

\begin{document}

\acmBooktitle{arXiv}

\title{State surveillance in the digital age: Factors associated with citizens' attitudes towards trust registers}
\renewcommand{\shorttitle}{Factors associated with citizens' attitudes towards trust registers}

\author{Katja Turha}
\email{katja.turha@student.um.si}
\affiliation{
  \institution{University of Maribor}
  \city{Maribor}
  \country{Slovenia}
}
\author{Simon Vrhovec}
\email{simon.vrhovec@um.si}
\affiliation{
  \institution{University of Maribor}
  \city{Maribor}
  \country{Slovenia}
}
\author{Igor Bernik}
\email{igor.bernik@um.si}
\affiliation{
  \institution{University of Maribor}
  \city{Maribor}
  \country{Slovenia}
}
\renewcommand{\shortauthors}{Turha et al.}

\begin{abstract}
This paper investigates factors related to the acceptance of trust registers (e.g., the Chinese Social Credit System -- SCS) in Western settings. To avoid a negative connotation, we first define the concept of trust register which encompasses surveillance systems in other settings beyond China, such as FICO in the US. Then, we explore which factors are associated with people's attitude towards trust registers leaning on the technology acceptance and privacy concern theories. A cross-sectional survey among Slovenian Facebook and Instagram users ($N=147$) was conducted. Covariance-based structural equation modeling (CB-SEM) was used to test the hypothesized associations between the studied constructs. Results indicate that attitude towards trust register is directly associated with perceived general usefulness of the trust register. Additionally, perceived general usefulness is associated with perceived usefulness of the trust register for ensuring national security and fighting crime, its ease of use, and privacy concern regarding data collection. As one of the first studies investigating attitude towards trust registers in a Western country, it provides pioneering insights into factors that may be relevant in case such registers would be implemented in a Western context, and provides some practical implications regarding messaging for would-be implementers of such systems.
\end{abstract}


\maketitle

\section{Introduction}
\label{section:intro}

The Social Credit System (SCS) developed in China is a widely discussed topic in the global community \citep{Vasilyeva2020}. Human rights advocates, government officials, and scholars of various specializations have taken an increasing interest in this phenomenon \citep{Bach2020,Vasilyeva2020,Kostka2019,Ma2019,Rieger2020,Mac2019,Ahmed2017,Ohlberg2017}. However, despite a high public profile, this problem is scientifically still poorly understood \citep{Vasilyeva2020}. 

The emerging SCS in China stands out as an initiative to radically transform society and the economy in the country \citep{Bach2020}, with a desire to take even greater control of its population and other entities. The system is envisioned to rate individuals, businesses, social organizations, and government agencies based on their level of \textit{trustworthiness}, and aims to be administered through various systems of punishments and rewards \citep{Kostka2019,Lin2023}. One of the more controversial features of this system is that it requires large amounts of personal data and information on each individual to function, which is collected from a variety of sources, such as financial, criminal, and government records, as well as various data from registry and school offices \citep{Backer2019,Ma2019,Li2023}. Notably, it also tracks subjects' activities online by including various data from digital sources \citep{Backer2019,Ma2019}. Digital data includes information collected on the internet, such as a person's online search history, shopping preferences, and social media interactions \citep{Ma2019}, control of which represents an intrusion into an individual's privacy \citep{Li2022}. In the future, the system could also include video system information obtained with facial recognition technology which is already widespread in China \citep{Backer2019} coupled with techniques, for example, for image superresolution (i.e., the process of enlarging and enhancing low-resolution images) \citep{Kumar2022} and age estimation \citep{Grd2023}, applications of artificial intelligence \citep{Zekan2022}, integration with VoIP mass surveillance systems \citep{Mathov2022}, or data obtained from smart systems, such as smart transportation \citep{Benyahya2022}. The development of the SCS itself is still ongoing so it cannot exactly be described as a single system \citep{Vasilyeva2020,Kostka2019,VonBlomberg2023ci}. For the time being, it is only an attempt to bring together different national, provincial, and municipal testing systems which focus on entirely different policies and issues  \citep{Backer2019,Wang2023}. This means that they do not have the same goal therefore the system cannot yet be unified. However, what all different SCS initiatives have in common is that they all seek to control through the establishment of a distributed state surveillance system fit for the current digital age and facilitate the rise of the digital society and cybernetic citizenship \citep{Hansen2023,Reijers2023,TrauthGoik2023}.

Such systems of control are not specific only for China, though \citep{Wang2023}. Different systems of control exist in other countries, such as Germany and the US, too, although that does not mean that they are socially recognized as such as well since they focus on the financial aspect \citep{Yin2023}. Schufa in Germany and EDGAR in the US are two examples of nationwide public platforms disclosing corporate information on listed companies \citep{Krause2023}. Among the best-known systems for individuals is the US FICO which indicates the creditworthiness of a particular individual \citep{Ignatius2018} and is used by 75 percent of lenders in the US \citep{Hendricks2011}. It was developed around 1950 \citep{Doroghazi2020} and officially came into use in 1989 \citep{Ignatius2018}. It is a system that evaluates whether or not an individual is creditworthy based on a credit score \citep{Hendricks2011, Ignatius2018, Arya2013}. According to this system, the higher the score, the higher the level of creditworthiness \citep{Brevoort2016}. The system uses five different factors \citep{Chatterjee2007, Demyanyk2010}, but the exact formula for calculating an individual's creditworthiness is not publicly known \citep{Hendricks2011, Arya2013, Demyanyk2010}. Banks use such systems to protect themselves from untrustworthy individuals making it easier for them to increase their profits \citep{Stewart2011}. A good way to predict an individual's actions is to evaluate their past \citep{Doroghazi2020, Demyanyk2010}. The credit score is based completely on the information contained in the credit report \citep{Hendricks2011} so the information obtained and processed must be correct. Numerous cases have shown that errors still occur in the processing of data (e.g., mistaken social security number, bankruptcy error) \citep{Hendricks2011,Brevoort2016} with individuals being severely affected as the errors impact their creditworthiness \citep{Arya2013}. To have a credit score, an individual must have at least one credit card or one loan, and if an individual does not have them, banks do not have enough information to calculate their credit score \citep{Elliott2018}. A 2016 survey found that approximately 45 million people in the US do not have a credit score \citep{Brevoort2016}. The consequences of a poor or undefined credit score significantly impact an individual's life and functioning in society, as they consequently have limited access to financial assistance \citep{Elliott2018, Brevoort2016}. Even when borrowing money, they are charged the highest possible interest rates which has a direct impact on their lives and their ability to save and improve their living situation \citep{Elliott2018, Brevoort2016}.

Both the Chinese SCS and comparable systems in other countries measure the level of trust of an individual and thus influence the individual's social functioning. To the best of our knowledge, the general concept of such surveillance systems has not been defined before. For the purpose of our study, we define the \textit{trust register} as an official register that can be introduced at the state level to monitor, assess and regulate the financial, social, moral and political behavior of natural and legal persons through a system of penalties and rewards. Trust registers provide various benefits to trusted people (e.g., tax breaks, easier access to loans and housing, cheaper public transport, shorter waiting times for health-related services, the possibility of renting a car without providing a security deposit) and through various penalties (e.g., tougher access to loans, limited access to public services, prohibition to perform state jobs, harder access to education) to encourage untrustworthy people to improve \citep{VonBlomberg2023jcc}. Trust registers draw information from a variety of traditional (e.g., financial, criminal, and state records, registry and school office data) and digital sources (e.g., online search history, online shopping history, social media activities) \citep{Li2023}. The data is accumulated into trust registers and processed automatically \citep{Yin2023}. The use of trust registers may be possible through a mobile application that allows individuals and organizations to see the level of trustworthiness of others, for example, online stores and potential customers. Everyone may quickly find out whether or how he wants to establish contact or do business with natural and legal persons whom he does not know yet or has no experience with \citep{VonBlomberg2023mc}.

A few studies have explored public opinion on SCS in China by examining citizens' attitudes toward the system and their privacy concerns \citep{Kostka2019,Rieger2020,Ahmed2017,Ohlberg2017}. Published literature suggests that SCS receives high levels of support among Chinese citizens which could be attributed to the lack of knowledge about the system \citep{Xu2023}. Studies also indicate that support is correlated with one's generalized fear \citep{Zeng2023}. On the organizational level, the use of SCS has also been associated with innovation \citep{Zuo2023}. However, there is a general lack of research that would investigate individuals' attitudes towards such surveillance systems outside of China. We found a single study, conducted in China, that indicates that public support for SCS may be lower when exposed to Western framing albeit only when individuals are informed about the monitoring of social behavior by SCS \citep{Xu2023}. Therefore, it might be safe to assume that such surveillance systems may not be as well-received in Western countries even though comparable systems are already in place there. To the best of our knowledge, there are no studies that would explore the factors associated with attitudes, adoption or rejection of trust registers outside of China.

This paper aims to address the above presented gaps in our understanding of what shapes one's attitude toward trust registers. This study makes four key contributions. First, this study defines the concept of trust register as an umbrella term for surveillance systems. Second, by considering trust registers as a technological innovation leveraged by a state for surveillance, our study leans on the technology acceptance theory to explore factors associated with attitude towards trust registers contributing to both state data surveillance and technology acceptance theory. Third, it is among the first to study how online privacy concerns are related to attitude towards trust registers contributing to the privacy concerns literature. Fourth, this study provides some insights into acceptance factors for implementations of trust registers with broadened scopes in Western countries.

\section{Research model}
\label{section:model}

\begin{figure}[!ht]
    \centering
    \includegraphics[width=.90\textwidth]{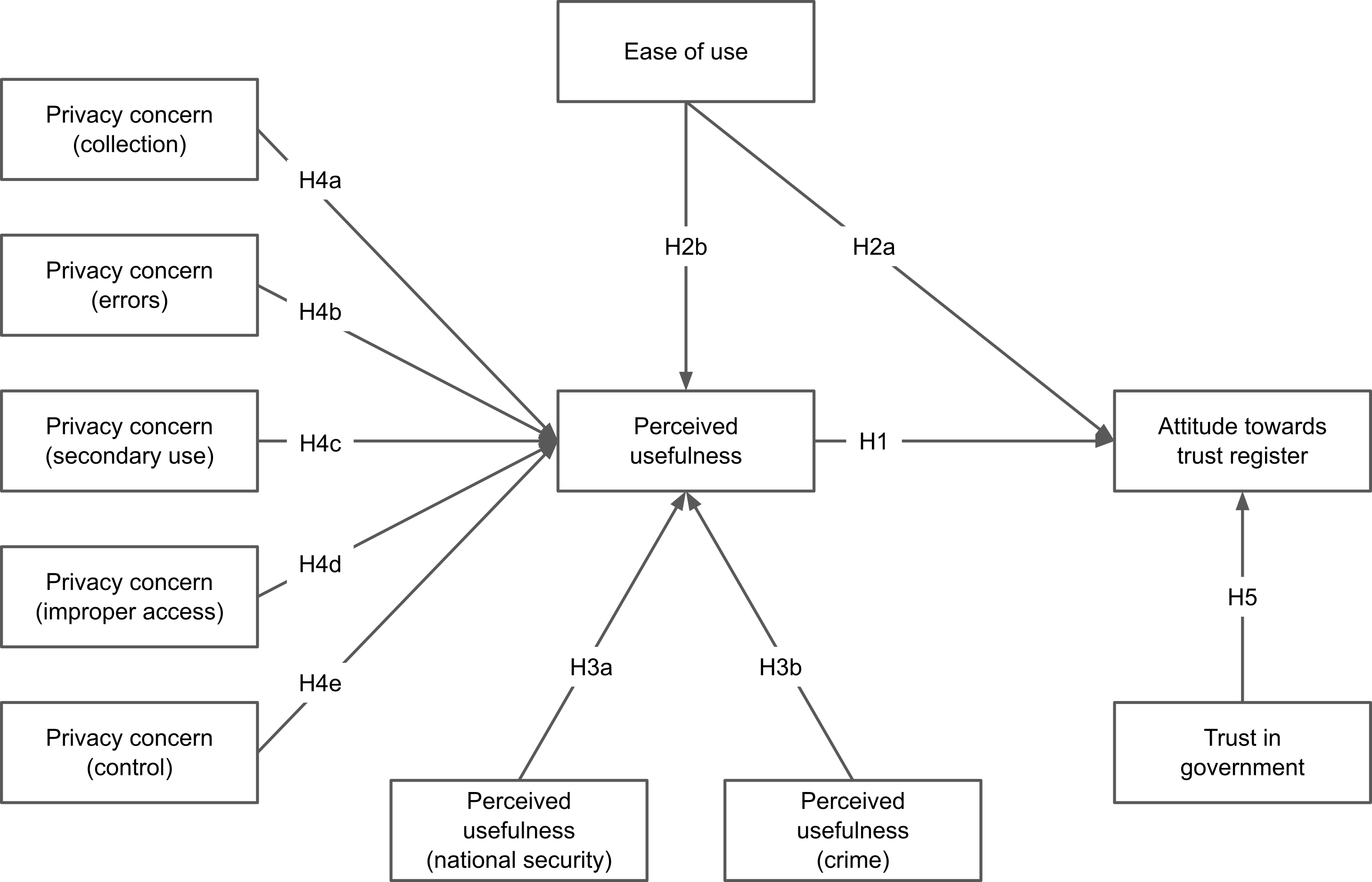}
    \caption{\label{fig:rm}Research model.}
\end{figure}

The study aims to investigate the main factors that would positively or negatively influence attitude toward trust registers. For this research, we have built a research model based on the hypotheses we have put forward. The model shows us the factors that we assume will have the most significant relations with attitude towards trust registers. The research model is presented in Figure~\ref{fig:rm}.

How individuals would accept the use of new technology can be tested using the Technology Acceptance Model (TAM) \citep{Davis1989}. The original TAM is based on two main predictors (i.e., perceived usefulness and perceived ease of use) of attitude towards use of new technology \citep{Davis1989}. Perceived usefulness measures the extent to which a person believes that using a new technology will increase their performance and gain benefits \citep{Venkatesh2000}. Perceived ease of use however helps us to determine the extent to which a person believes that they would be able to use the new technology effortlessly and that it would be easy to use \citep{Davis1989}. In other words, the more benefits the user gets and the easier new technology is to use, the more likely individuals will support it and be willing to use it \citep{Alassafi2022}. Based on the original TAM model, we therefore formulate the following hypotheses:

\begin{itemize}[leftmargin=1.4cm]
    \item[\textit{H1}:]{Perceived usefulness of the trust register is positively associated with attitude towards trust register.}
    \item[\textit{H2a}:]{Ease of use of the trust register is positively associated with attitude towards trust register.}
    \item[\textit{H2b}:]{Ease of use of the trust register is positively associated with perceived usefulness of the trust register.}
\end{itemize}

Perceived usefulness can be measured by various indicators, such as speed of achievement, increased productivity, efficiency, etc. \citep{Warkentin2007}. In the context of trust registers, there are two key benefits that are emphasized by their implementers, namely ensuring national security and fighting crime. For the purpose of our study, we assume these two dimensions of perceived usefulness shape the perceived general usefulness of trust registers. Based on these considerations, we develop the following set of hypotheses:

\begin{itemize}[leftmargin=1.4cm]
    \item[\textit{H3a}:]{Perceived usefulness of the trust register for ensuring national security is positively associated with perceived general usefulness of the trust register.}
    \item[\textit{H3b}:]{Perceived usefulness of the trust register for fighting crime is positively associated with perceived general usefulness of the trust register.}
\end{itemize}

Information privacy pertains to an individual's capacity to personally manage the information concerning their identity \citep{Stone1983}. When discussing measuring privacy concerns, we can divide them into five different dimensions \citep{Smith1996,VanSlyke2006,Fortes2016,Pu2022}. First, concern regarding data collection is connected with the collection and storage of personal data in a certain system. Second, concern regarding secondary use of data is about the use of stored data in a certain system for a different purpose than it was collected for. Third, concern regarding improper access is related to data stored in a certain system being accessed by unauthorized persons. Fourth, concern regarding errors is linked to errors in collected data stored in a certain system. Fifth, concern regarding control is connected with (the lack of) control that individuals have over their collected data stored in a certain system. People who worry more about their privacy may therefore feel mentally burdened when they need to share their personal data with trust registers decreasing their motivation to engage with such systems \citep{Davis1992,Fortes2016}. However, such feelings have a negative impact on perceived usefulness \citep{Fortes2016}, and individuals will be less likely to use technology that lowers their expectations of privacy \citep{Dhagarra2020}. Based on the privacy literature, we thus pose the following set of hypotheses:

\begin{itemize}[leftmargin=1.4cm]
    \item[\textit{H4a}:]{Privacy concern about data collection is negatively associated with perceived usefulness of the trust register.}
    \item[\textit{H4b}:]{Privacy concern about secondary use of data is negatively associated with perceived usefulness of the trust register.}
    \item[\textit{H4c}:]{Privacy concern about improper access to data is negatively associated with perceived usefulness of the trust register.}
    \item[\textit{H4d}:]{Privacy concern about data errors is negatively associated with perceived usefulness of the trust register.}
    \item[\textit{H4e}:]{Privacy concern about data control is negatively associated with perceived usefulness of the trust register.}
\end{itemize}

Trust may equal power and can therefore be valuable in many interactions. Establishing trust is a long process as it develops over a long period of time, but can be lost in an instant \citep{Tolbert2006}. Trust is a psychological state in which an individual is willing to acknowledge, accept, or show their vulnerability to a particular individual or the public at large because they expect positive intentions from the other party \citep{Cullen2008}. Trust also has a strong influence on the performance of governments \citep{Colesca2009}. Studies show that in modern democracies citizens' distrust of their government can have a negative impact on its performance \citep{Nasser2020}. Our trust in institutions is conditioned by our expectations, knowledge of their functioning and intentions, and the competence of the individuals who work within them \citep{Colesca2009}. The greater the trust in government, the more willing citizens are to engage with it \citep{Tolbert2006}. Citizen engagement with government is necessary because government institutions and agencies need personal data to operate, which they collect and process \citep{Nasser2020} to ensure the proper functioning of government systems such as e-government, healthcare, etc. Providing personal data to government organizations is therefore absolutely necessary in many cases \citep{Cullen2008}. So it makes it all the more important that government and government organizations handle sensitive information correctly and use the data only for the purposes for which it is collected \citep{Nasser2020}. Individuals who do not trust the government and its performance are skeptical about participating \citep{Cullen2008} in different surveillance systems, including trust registers. Based on this, we developed the final hypothesis:

\begin{itemize}[leftmargin=1.4cm]
    \item[\textit{H5}:]{Trust in government is negatively associated with attitude towards trust register.}
\end{itemize}

The research model therefore draws from the technology acceptance, privacy concern and trust theories, and tries to draw synergies in explaining how attitudes towards trust registers are formed.

\section{Methods}
\label{section:method}

\subsection{Research design}

Our study used a cross-sectional research design to investigate how the Slovenian population would accept a trust register (i.e., a surveillance system comparable to the SCS). In order to identify the factors related to attitude towards a trust register, we conducted the survey among Slovenian Facebook and Instagram users.

\subsection{Ethical considerations}

Approval from the Institutional Review Board for this study was not required according to the legislation of the Republic of Slovenia and internal acts of the University of Maribor.

\subsection{Measures}

The questionnaire measured 12 theoretical constructs shown in Table~\ref{table:constructs}. We measured the following constructs: attitude towards trust register, perceived usefulness (overall, national security, crime), ease of use, privacy concern (collection, errors, secondary use, improper access, control) and trust in government. For the purposes of our research, we have took or adapted previously validated construct items to the context of our study. We took the items for trust in government from \citep{Jelovcan2021}. Items for all privacy concern constructs were adapted from \citep{Hong2013}. Items for perceived usefulness were adapted from \citep{Ma2017}. Items for perceived usefulness (national security), perceived usefulness (crime) and ease of use were adapted from \citep{Venkatesh2003}. Items for attitude towards trust register were adapted from \citep{Siegel2014}. Items were measured using Likert and bipolar scales. Items for attitude towards the trust register were measured with a 5-point bipolar scale. Items for all privacy concern constructs were measured by using a 5-point Likert scale from 1 (\textit{strongly disagree}) to 5 (\textit{strongly agree}). The remaining items were measured with a 7-point Likert scale from 1 (\textit{strongly disagree}) to 7 (\textit{strongly agree}). The survey was conducted in the Slovenian language.

\begin{table}[!ht]
\footnotesize
\caption{\label{table:constructs}Theoretical construct definitions.}
\begin{tabular}{ll}
\toprule
Theoretical construct & Operational definition \\
\midrule
Attitude towards trust register & An individual’s positive versus negative evaluations of the trust register. \\
Perceived usefulness & The perceived general usefulness of the trust register. \\
Perceived usefulness (national security) & The perceived usefulness of the trust register for ensuring national security. \\
Perceived usefulness (crime) & The perceived usefulness of the trust register for fighting crime.  \\
Ease of use & The perceived ease of use of the trust register. \\
Privacy concern (collection) & The extent of concerns regarding the collection of personal information by the trust register. \\
Privacy concern (errors) & The extent of concerns regarding errors in personal information stored in the trust register. \\
Privacy concern (secondary use) & The extent of concerns regarding secondary use of personal information stored in the trust register. \\
Privacy concern (improper access) & The extent of concerns regarding improper access to personal information stored in the trust register. \\
Privacy concern (control) & The extent of concerns regarding control over personal information stored in the trust register. \\
Trust in government & The extent of trusting beliefs in the government. \\
\bottomrule
\end{tabular}
\end{table}

\subsection{Data collection}

The survey was available from August to December 2021. The invitation to take the survey was posted in 69 Facebook groups and shared 11 times on Instagram, implying a convenience sample. Participation in our survey was voluntary and anonymous. 155 respondents participated in our survey. We excluded seven responses with over 10 percent missing values and one response indicating respondent non-engagement. We ended up with $N=147$ usable responses for the analysis. Characteristics of the sample are presented in Table~\ref{table:sample}. The age of respondents ranged from 17 to 69 years old ($M=32.1,SD=12.0$). The primary source of information for 77.6 percent of respondents was the internet indicating a sample biased towards internet users. This is likely a consequence of conducting the survey as an online questionnaire.

\begin{table}[!ht]
\footnotesize
\caption{\label{table:sample}Sample characteristics.}
\begin{tabular}{llrr}
\toprule
Characteristic & & Frequency & Percent \\
\midrule
Gender & Female & $105$ & $71.4$ \\
 & Male & $41$ & $27.9$ \\
 & N/A & $1$ & $0.7$ \\
Employment status & Student & $47$ & $32.0$ \\
 & Employed / Self-employed & $82$ & $55.8$ \\
 & Farmer / Housewife & $2$ & $1.4$ \\
 & Unemployed & $9$ & $6.1$ \\
 & Retired & $7$ & $4.8$ \\
Formal education & Finished high school or less & $58$ & $39.5$ \\
 & Acquired Bachelor's degree & $54$ & $36.7$ \\
 & Acquired Master's degree  & $28$ & $19.0$ \\
 & Acquired PhD degree  & $7$ & $4.8$ \\
Living environment & Urban & $91$ & $61.9$ \\
 & Rural & $56$ & $38.1$ \\
Status & Single & $51$ & $34.7$ \\
 & In a relationship -- not living together & $19$ & $12.9$ \\
 & In a relationship -- living together & $38$ & $25.9$ \\
 & Married & $35$ & $23.8$ \\
 & Divorced & $4$ & $2.7$ \\
Primary source of information & Printed media & $2$ & $1.4$ \\
 & Radio & $7$ & $4.8$ \\
 & TV & $15$ & $10.2$ \\
 & Internet & $114$ & $77.6$ \\
 & Family and friends & $9$ & $6.1$ \\
\bottomrule
\end{tabular}
\end{table}

\subsection{Data analysis}

To analyze the data, we employed covariance-based structural equation modeling (CB-SEM). The key advantage of this data analysis method is that it integrates into a concurrent evaluation latent variables with multiple indicators and their inter-relations. We used R ver. 4.3.1, lavaan ver. 0.6–15 and semTools ver. 0.5-6 to analyze the data. Prior to the analysis, we imputed missing values (0.2 percent) with medians. We used standard model fit indices and thresholds: $\chi^2 / df$ -- $< 2.0$ excellent fit, $2.0-5.0$ good fit, $> 5.0$ poor fit; CFI -- $> 0.95$ excellent fit, $0.90-0.95$ good fit, $< 90.0$ poor fit; TLI -- $> 0.95$ excellent fit, $0.90-0.95$ good fit, $< 90.0$ poor fit; RMSEA -- $< 0.06$ excellent fit, $0.06-0.10$ good fit, $> 0.10$ poor fit; and SRMR -- $< 0.06$ excellent fit, $0.06-0.08$ good fit, $> 0.08$ poor fit. We conducted a confirmatory factor analysis (CFA) to validate the survey instrument. First, we determined convergent validity by evaluating AVE and factor loadings of questionnaire items. AVE values above the $0.50$ threshold are considered acceptable. Next, we evaluated discriminant validity with a HTMT analysis. Ratios of correlations below the $0.90$ threshold are considered acceptable. Finally, we determined reliability with CR and CA. CR and CA values above the $0.70$ threshold are considered acceptable. A structural model was constructed to test the hypothesized associations.

\section{Results}
\label{section:results}

\subsection{Instrument validation}

We first developed a measurement model to validate the measurement instrument. Model fit of the measurement instrument is presented in Table~\ref{table:mm_fit}. It shows that the data fits the model well.

\begin{table}[!ht]
\footnotesize
\caption{\label{table:mm_fit}Fit indices of the measurement model.}
\begin{tabular}{lcrl}
\toprule
Measure & Threshold & Estimate & Interpretation \\
\midrule
$\chi^2$ & & $641.186$ & \\
$df$ & & $440$ & \\
$\chi^2 / df$ & $\leq 5$ & $1.457$ & Excellent \\
CFI & $\geq 0.90$ & $0.965$ & Excellent \\
TLI & $\geq 0.90$ & $0.959$ & Excellent \\
RMSEA & $\leq 0.08$ & $0.056$ & Excellent \\
SRMR & $\leq 0.08$ & $0.040$ & Excellent \\
\bottomrule
\end{tabular}
\begin{flushleft}
\textit{Notes}: CFI -- comparative fit index; TLI -- Tucker-Lewis index; RMSEA -- root mean square error of approximation; SRMR -- standardized root mean square residual.
\end{flushleft}
\end{table}

Table~\ref{table:validity} presents the results of analyses relevant for determining the validity and reliability of the survey instrument. First, CA ranged from 0.859 to 0.981 and CR ranged from 0.861 to 0.981 demonstrating adequate reliability of all constructs. Second, AVE ranged from 0.676 to 0.945 indicating adequate convergent validity. Third, HTMT analysis indicates that discriminant validity of the measurement instrument is adequate.

\begin{table*}[!ht]
\footnotesize
\caption{\label{table:validity}Survey instrument validation. Cronbach's alpha (CA), composite reliability (CR), average variance extracted (AVE), and heterotrait-monotrait ratio of correlations (HTMT) analysis.}
\begin{tabular}{lrrrrrrrrrrrrr}
\toprule
Construct & CA & CR & AVE & 1 & 2 & 3 & 4 & 5 & 6 & 7 & 8 & 9 & 10 \\
\midrule
1: PCc & 0.910 & 0.908 & 0.768 &  &  &  &  &  &  &  &  &  &  \\
2: PCsu & 0.957 & 0.958 & 0.885 & 0.761 &  &  &  &  &  &  &  &  &  \\
3: PCe & 0.926 & 0.925 & 0.805 & 0.541 & 0.593 &  &  &  &  &  &  &  &  \\
4: PCia & 0.957 & 0.958 & 0.884 & 0.766 & 0.832 & 0.646 &  &  &  &  &  &  &  \\
5: PCctl & 0.937 & 0.937 & 0.833 & 0.759 & 0.852 & 0.609 & 0.767 &  &  &  &  &  &  \\
6: TiG & 0.859 & 0.861 & 0.676 & 0.250 & 0.284 & 0.314 & 0.289 & 0.338 &  &  &  &  &  \\
7: PU & 0.953 & 0.954 & 0.873 & 0.128 & 0.131 & 0.047 & 0.220 & 0.038 & 0.105 &  &  &  &  \\
8: PUns & 0.981 & 0.981 & 0.945 & 0.050 & 0.023 & 0.090 & 0.073 & 0.039 & 0.106 & 0.828 &  &  &  \\
9: PUc & 0.971 & 0.971 & 0.919 & 0.070 & 0.017 & 0.074 & 0.089 & 0.041 & 0.155 & 0.759 & 0.764 &  &  \\
10: EoU & 0.888 & 0.887 & 0.725 & 0.054 & 0.118 & 0.092 & 0.168 & 0.019 & 0.101 & 0.689 & 0.553 & 0.549 &  \\
11: AtTR & 0.891 & 0.890 & 0.729 & 0.095 & 0.200 & 0.028 & 0.247 & 0.111 & 0.091 & 0.864 & 0.777 & 0.685 & 0.612 \\
\bottomrule
\end{tabular}
\begin{flushleft}
\textit{Notes}: PCc -- privacy concern (collection); PCsu -- privacy concern (secondary use); PCe -- privacy concern (errors); PCia -- privacy concern (improper access); PCctl -- privacy concern (control); TiG -- trust in government; PU -- perceived usefulness; PUns -- perceived usefulness (national security); PUc -- perceived usefulness (crime); EoU -- ease of use; AtTR -- attitude towards trust register.
\end{flushleft}
\end{table*}

\begin{table*}[!ht]
\scriptsize
\caption{\label{table:items}Questionnaire items.}
\begin{tabular}{lrp{.60\textwidth}l}
\toprule
Construct & Loading & Prompt / Item & Source \\
\midrule
Privacy concerns (collection) & 0.865 & PCc1. It usually bothers me when e-government websites ask me for personal information. & \citep{Hong2013} \\
& 0.848 & PCc2. When e-government websites ask me for personal information, I sometimes think twice before providing it. &  \\
& 0.911 & PCc3. I am concerned that e-government websites are collecting too much personal information about me.  &  \\
Privacy concerns (secondary usage) & 0.909 & PCsu1. I am concerned that when I give personal information to a e-government website for some reason, the website would use the information for other reasons. & \citep{Hong2013} \\
& 0.964 & PCsu2. I am concerned that e-government websites would sell my personal information in their computer databases to public or private companies. &  \\
& 0.946 & PCsu3. I am concerned that e-government websites would share my personal information with public or private companies without my authorization. &  \\
Privacy concerns (errors) & 0.909 & PCe1. I am concerned that e-government websites do not take enough steps to make sure that my personal information in their files is accurate. & \citep{Hong2013} \\
& 0.874 & PCe2. I am concerned that e-government websites do not have adequate procedures to correct errors in my personal information. &  \\
& 0.910 & PCe3. I am concerned that e-government websites do not devote enough effort to verifying the accuracy of my personal information in their databases. &  \\
Privacy concerns (improper access) & 0.916 & PCia1. I am concerned that e-government website databases that contain my personal information are not protected from unauthorized access. & \citep{Hong2013} \\
& 0.956 & PCia2. I am concerned that e-government websites do not devote enough effort to preventing unauthorized access to my personal information. &  \\
& 0.950 & PCia3. I am concerned that e-government websites do not take enough steps to make sure that unauthorized people cannot access my personal information. &  \\
Privacy concerns (control) & 0.920 & PCctl1. It usually bothers me when I do not have control of personal information that I provide to e-government. & \citep{Hong2013} \\
& 0.914 & PCctl2. It usually bothers me when I do not have control over decisions about how my personal information is collected, used, and shared by e-government websites. &  \\
& 0.905 & PCctl3. I am concerned when control over my personal information is lost as a result of a marketing transaction with e-government websites. &  \\
Trust in government & 0.933 & TiG1. I believe that the government would act in my best interest. & \citep{Jelovcan2021} \\
& 0.641 & TiG2. The government is interested in my well-being not just its own. &  \\
& 0.933 & TiG3. I would characterize the government as honest. &  \\
Perceived usefulness & 0.954 & PU1. Being included in a trust register would be useful for me. & \citep{Ma2017} \\
& 0.963 & PU2. Being included in a trust register would be very beneficial for me. &  \\
& 0.890 & PU3. Being included in a trust register would give me access to useful information. &  \\
Perceived usefulness (national security) & 0.966 & PUns1. A trust register would make it easier to ensure national security. & \citep{Venkatesh2003} \\
& 0.989 & PUns2. I would find a trust register useful in ensuring national security. &  \\
& 0.962 & PUns3. A trust register would enhance the effectiveness of ensuring national security. &  \\
Perceived usefulness (crime) & 0.952 & PUc1. A trust register would make it easier to fight crime. & \citep{Venkatesh2003} \\
& 0.944 & PUc2. I would find a trust register useful in fighting crime. &  \\
& 0.979 & PUc3. A trust register would enhance the effectiveness of fighting crime. &  \\
Ease of use & 0.903 & EoU1. My interaction with a trust register would be clear and understandable. & \citep{Venkatesh2003} \\
& 0.771 & EoU2. It would be easy for me to become skillful at using a trust register. &  \\
& 0.876 & EoU3. I would find a trust register easy to use. &  \\
Attitude towards trust register &  & In general, how do you feel about trust registers? & \citep{Siegel2014} \\
& 0.840 & AtTR1. Negative ... Positive &  \\
& 0.891 & AtTR2. Undesirable ... Desirable &  \\
& 0.828 & AtTR3. Harmful ... Beneficial &  \\
\bottomrule
\end{tabular}
\end{table*}

\clearpage

\subsection{Structural model}

We developed a structural model to test the hypothesized associations. Model fit of the structural model is presented in Table~\ref{table:sm_fit}. It indicates that the model fits the data well. 

\begin{table}[!ht]
\footnotesize
\caption{\label{table:sm_fit}Fit indices of the structural model.}
\begin{tabular}{lcrl}
\toprule
Measure & Threshold & Estimate & Interpretation \\
\midrule
$\chi^2$ & & $650.845$ & \\
$df$ & & $448$ & \\
$\chi^2 / df$ & $\leq 5$ & $1.453$ & Excellent \\
CFI & $\geq 0.90$ & $0.965$ & Excellent \\
TLI & $\geq 0.90$ & $0.959$ & Excellent \\
RMSEA & $\leq 0.08$ & $0.055$ & Excellent \\
SRMR & $\leq 0.08$ & $0.041$ & Excellent \\
\bottomrule
\end{tabular}
\begin{flushleft}
\textit{Notes}: CFI -- comparative fit index; TLI -- Tucker-Lewis index; RMSEA -- root mean square error of approximation; SRMR -- standardized root mean square residual.
\end{flushleft}
\end{table}

\begin{figure}[!ht]
    \centering
    \includegraphics[width=.90\textwidth]{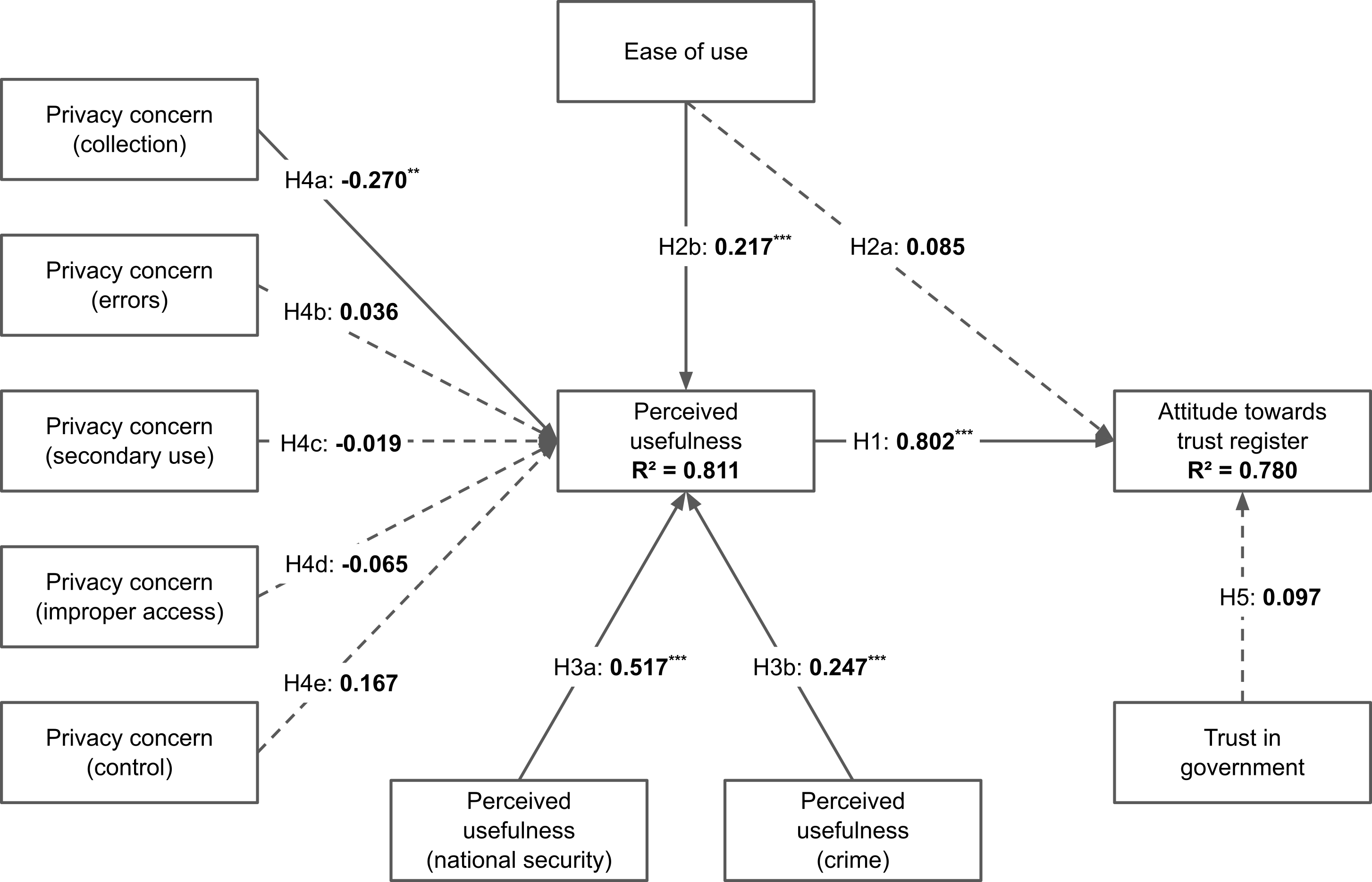}
    \caption{\label{fig:sm}Structural model.}
\end{figure}

Standardized results of the structural model are presented in Figure~\ref{fig:sm}. The results of the structural model support hypotheses H1 ($p<0.001$), H2b ($p<0.001$), H3a ($p<0.001$), H3b ($p<0.001$) and H4a ($p=0.003$). However, the results do not indicate support for hypotheses H2a ($p=0.260$), H4b ($p=0.558$), H4c ($p=0.860$), H4d ($p=0.487$), H4e ($p=0.089$) and H5 ($p=0.065$).

\section{Discussion}
\label{section:discussion}

This study is among the first to investigate factors associated with attitude towards trust registers by leaning on the technology acceptance and privacy concerns theory. It is also among the first such studies focusing on a population outside China. It makes a number of contributions to the literature on attitudes toward state data surveillance, technology acceptance and privacy concerns, and provides some practical implications regarding messaging for would-be implementers of such systems in Western contexts. First, this study defines the concept of trust register as an umbrella term for surveillance systems comparable to the SCS. SCS and comparable systems in the West are essentially registers in which data are accumulated and processed to determine an individual's or other entity's trustworthiness score. The aim of defining a new neutral concept instead of focusing the study on SCS was to avoid the effect of the potential negative connotation that SCS may have in Western countries, such as Slovenia. Although studies focusing on SCS may have some merit in Western countries too, the negative connotation due to the Western media exposure and the fact that SCS originates from China may affect the validity of such studies. Therefore, the implications of such studies might have little relevance for potential implementations of SCS-like systems in Western countries. Since trust registers encompass both SCS and other surveillance systems, using this concept may enable researchers to study the impact of such (non-)negative connotations of both SCS and other comparable systems.

Second, this is one of the first studies to investigate which factors are associated with attitude towards trust register. The results of our study indicate that the associations based on the original TAM have merit in the context of our study. More specifically, perceived usefulness is directly associated with attitude towards trust registers. Even though ease of use was not directly related to attitude towards trust register as predicted by the original TAM, there is an indirect relation mediated by perceived usefulness as predicted by the original TAM \citep{Marangunic2015}. We additionally found that perceived general usefulness is associated with both perceived usefulness of the trust register for ensuring national security and fighting crime. These findings indicate that these often emphasized SCS goals are indeed relevant in shaping the attitude towards trust register and thus effective in promoting SCS to the public. Although trust registers are enabled and may be ultimately controlled by governments, trust in government was not associated with attitude towards trust register. This may be a consequence of the study settings since it was conducted in a Western democracy as countries vary considerably according to their residents' surveillance concern, fear of government intrusions into privacy and trust in government \citep{Fujs2019}. Future studies in other countries around the globe (e.g., autocratic regimes) would be necessary to determine whether there are significant differences in other political contexts.

Third, the results of our study indicate that privacy concern about data collection is indirectly associated with attitude towards trust register through perceived usefulness. Although the published literature emphasizes privacy concern of citizens due to the overreaching nature of trust registers \citep{Kostka2019,Rieger2020,Ahmed2017,Ohlberg2017}, this is one of the first studies that empirically test these assumptions. The results of this study suggest that only one out of five tested dimensions of privacy concern is associated with attitude towards trust register. These findings is therefore only partially in line with the published SCS literature as respondents seem to associate only data collection with their attitude toward trust register, and not other privacy-related issues, such as errors in collected data, secondary use of the data, improper access to the data (e.g., by hackers) or the lack of control over collected data. Nevertheless, this may be yet another example of the privacy paradox according to which people engage in privacy infringing behavior despite voicing concern about it \citep{Lenz2023}.

Fourth, the results of this study provide some insights into the possible implementation of trust registers in Western countries. As already noted, trust registers are not unheard of in these countries (e.g., EDGAR, Schufa, FICO \citep{Krause2023,Ignatius2018}). However, these systems focus solely on the financial aspect \citep{Yin2023}. Even though Western trust registers are not associated with the social aspect of surveillance, surveillance of social media data is extensive in Western countries too \citep{Schyff2020,Selvarajah2022}. The results of our study provide some insights into acceptance factors for potential implementers of trust registers with broadened scopes (e.g., the social aspect) in Western countries. Messaging regarding the implementation of a trust register with a broadened scope may focus on its perceived usefulness. The results of our study suggest that perceived general usefulness of trust registers may be shaped by its ease of use and usefulness for ensuring national security and fighting crime. Key messaging points could therefore focus on these topics to shape an adequate attitude towards trust register. These topics are already ingrained in Western societies as acceptable reasons for exchanging privacy for security \citep{DaSilva2022}. Additionally, the results of our study indicate that messaging aiming to relieve the privacy concern may primarily focus on alleviating people's concern regarding the collection of data. These practical implication should however be taken with some reserve as it assumes that the target population does not associate such a trust register with SCS or similar systems. Should the population relate the implementation of a trust register to the implementation of a controversial surveillance system, the negative connotation of such a trust register might significantly alter acceptance factors. Future studies, such as those focusing on a potential implementation of a SCS-like surveillance system in the West, may be needed to estimate the extent of these changes.

This study has some limitations that the readers should note. First, the sample is not representative therefore the readers should be careful when generalizing the results to the studied population. Future studies may aim to include respondents that were underrepresented in our sample, notably people whose primary source of information is not the internet. Data collection methods beyond online surveys may be employed to achieve these aims. Second, the study was conducted in a single Western country. For improving the ecological validity of the study, future studies may focus on other countries in the Western cultural context, too. To better understand the differences between the Eastern and Western cultural contexts as well as in countries with varying surveillance concerns \citep{Fujs2019}, comparative studies would be highly beneficial as well. Third, the study described trust registers to the respondents without providing them with any real-world examples of such registers. Although the description was based on the SCS, the respondents may not have grasped all implications of implementing trust registers on a large scale. Future studies focusing on existing trust registers, such as SCS, EDGAR, Schufa or FICO, or implementations of other pilot trust registers, perhaps for the sole purpose of conducting experiments, may thus be beneficial. Fourth, this study focused on attitude towards trust register. Attitude is an important albeit not the only acceptance factor. Future studies may include other acceptance factors, such as social influence, in their investigations. However, the role of some of these factors may be hard to study without actual trust register implementations. Additionally, studying factors associated with actual acceptance in pilot implementations of trust registers would be highly beneficial. Finally, our study focused on individuals. Trust registers can be used for both individuals and organizations. Therefore, future studies may explore the factors associated with acceptance of trust registers by organizations.

\begin{acks}
\small
The authors thank the respondents for taking their time to participate in the study.
\end{acks}

\bibliographystyle{ACM-Reference-Format}


\end{document}